# Managing Requirements Change the Informal Way:
## When Saying 'No' is Not an Option

Waqar Hussain[1], Didar Zowghi[2], Tony Clear[3], Stephen MacDonell[4], Kelly Blincoe[5]
[1,3,4]School of Computer and Mathematical Sciences, Auckland University of Technology, Auckland, New Zealand
[2]Faculty of Engineering and Information Technology, University of Technology Sydney, Australia
[5]Department of Electrical and Computer Engineering, University of Auckland, New Zealand
[1,3,4]{whussain, tclear, smacdone} @aut.ac.nz, [2]didar.zowghi@uts.edu.au, [5]kblincoe@acm.org

**Abstract**

*Software has always been considered as malleable. Changes to software requirements are inevitable during the development process. Despite many software engineering advances over several decades, requirements changes are a source of project risk, particularly when businesses and technologies are evolving rapidly. Although effectively managing requirements changes is a critical aspect of software engineering, conceptions of requirements change in the literature and approaches to their management in practice still seem rudimentary.*

*The overall goal of this study is to better understand the process of requirements change management. We present findings from an exploratory case study of requirements change management in a globally distributed setting. In this context we noted a contrast with the traditional models of requirements change. In theory, change control policies and formal processes are considered as a natural strategy to deal with requirements changes. Yet we observed that "informal requirements changes" (InfRc) were pervasive and unavoidable. Our results reveal an equally 'natural' informal change management process that is required to handle InfRc in parallel. We present a novel model of requirements change which, we argue, better represents the phenomenon and more realistically incorporates both the informal and formal types of change.*

## 1. INTRODUCTION

Requirements change is a recognized and accepted phenomenon in contemporary software development. In fact, change is welcomed and embraced in agile development approaches, as a means of adding value and improving usability. On the other hand, uncontrolled changes may pose a risk to cost and quality of software [1, 2] and hurt organizations through missed deadlines, budget overruns and wasted resources [3].

Requirements evolve due to a combination of internal and external factors that trigger change [1]. Some of the unavoidable changes that impact the development process are still manageable because they are customer-initiated, externally focused and assessable in terms of their impact. Therefore communicating the implications of these changes to customers is comparatively straight forward and establishing change control policies or safeguards against them is possible [4, 5]. In this study we have observed another class of requirements change that we refer to as *Informal Requirements Change* (InfRC). InfRCs are more internally focused, potentially subversive to the development process and therefore harder to manage. We define InfRCs as changes in requirements initiated by any stakeholder that bypass most of the policies or controls imposed by formal change management processes (e.g. formal review and change impact analysis) and are implemented in the evolving system.

Several factors contribute to the manifestation of InfRC in software development projects. Sometimes they arise as a consequence of prematurely ending RE activities [6] or attempting a requirements 'freeze' earlier than usual in a project, thus 'latent' but necessary changes spring up [5]. In other cases InfRC might emerge as a consequence of skunkworks (work hidden by managers to get something developed by making ad hoc decisions and bypassing time consuming formalities [7]), creeping requirements (a continuous influx of requirements additions and changes) [8] or creeping elegance [9] (additions made without the consideration of delay in the schedule and project cost [10]). InfRCs may also result from the failure to create a practical process to help manage changes [11].

We posit that although many projects still use plan-driven RCM processes for good reasons but they seem to lack adequate support to recognize and manage InfRC. The overall objective of this research is thus to explore the notion of InfRC by conducting a case study. Our aim is to increase our understanding of this complex phenomenon, discover its sources, the reasons to accommodate them and the implications of dealing with such changes in an informal manner. The research questions for our study are:

*RQ1. What are the sources of informal changes to requirements?*

*RQ2. How are informal changes handled in practice?*

*RQ3. What are the implications of managing requirements changes informally?*

In addition to presenting the findings of our exploratory case study of managing InfRC in a software development project, this paper also presents a more realistic change management process model for InfRC. To our knowledge this is the first known model that captures both formal and informal activities to manage requirements. This paper is organized as follows; Section 2 briefly describes the unpredictable nature of requirements and presents the classical perspective on requirements change management. Section 3 describes the research methods used in this case study. Section 4 presents the research settings and profiles our case study adopting a vendor's perspective. Section 5 highlights the findings and presents the model of requirements change derived from this study, discusses the sources of InfRC, the reasons to accommodate changes informally and its implications. Section 6 reflects on InfRC as an inevitable phenomenon and the oversight with regards to InfRC in the existing RCM models in literature. The limitations of this research are covered next in Section 7, followed by the implication of the results for research in Section 8. Section 9 briefly concludes the paper.

## 2. BACKGROUND

Commercial bespoke projects continue to face an influx of requirements change from elicitation through to delivery and even beyond [1]. This reality shatters the rigid, and unnatural formal change control policies superimposed by management on the projects to keep them under control.

The reality of developing software is its innate malleability and the emerging (sometimes arbitrary) nature of requirements. The initial vision for a software solution evolves as the project is explored through dynamic artefacts that clarify the initial perception of reality [12]. Similarly stakeholders with different opinions and priorities express their requirements in different ways leading to ambiguities and inconsistencies [13]. Often changes in requirements need to be made to resolve them. Some of these changes are handled by a formal process while others follow an informal path.

The traditional RCM process models found in the literature are geared towards handling requirements change based on formal change control policies [5, 8, 14-16]. The drivers for these models appear to be both commercial as well as project management concerns of controlling cost and scope. The underlying assumptions in almost all models is that changes only occur when requirements are base-lined and therefore changes should only be treated formally. However in reality, the relationship between the change requestor and implementer and the urgency or significance of change may not allow a change to always follow a formal path for implementation [16, 17]. For example, prototypes can be informally "hacked together" by both the customer and the developers. When clients or their representatives have easy access to development teams they often request additions to the requirements without going through a formal change review process [9]. In such circumstances customers often approach developers directly to get their desired changes implemented into the system. Similarly the developers can (informally) add features of their own choice to the software by means of 'gold plating' [18]. Thus, requirements can become unstable in ways that are not always visible to project managers [9].

The existing requirements management process models do not acknowledge or treat such informal changes. Under conventional change management approaches, the prescribed measures to efficiently manage scope creep include having a single channel to handle change with a firewall to guard against unwanted changes [5], base-lining requirements [8] and checking, costing and approving changes. However, none of these approaches are specifically designed to handle informal changes in requirements as described previously.

We have identified a context wherein informal changes in requirements were inevitable and pervasive. A formal change process may appear a 'natural' strategy to cope with changes in requirements in a formal way. However, our study suggests that there is also an equally 'natural' and parallel process that occurs through which informal changes are handled.

## 3. RESEARCH METHODS

To answer our research questions, we performed a case study on a decision support software development project carried out across three geographically distributed client and vendor sites in the USA and Pakistan. An exploratory case study methodology [19] was applied to gain a deeper understanding of the phenomenon of requirements change, which we argue is an under-theorized area in software engineering. Data was collected primarily from semi-structured interviews, observation of the requirements management process, and inspection of change related artefacts (e.g. RM process documentation, Requirements Change and Issue Logs).

**A. Interviews**
The first author travelled to Pakistan to carry out interviews of the key project stakeholders from the vendor side. Seventeen semi-structured interviews of approximately 45 minutes each were conducted in two phases for the case study. In phase two findings from initial analysis were confirmed and follow up questions were asked to identify the evolution of practices (if any) in managing requirements changes. The interviewees included one development manager, two team leads, two developers and a quality assurance manager. To cover the client's perspective, the CEO of the company was interviewed who also acted as the proxy to the client. The interviews were guided by high level questions such as: "what are the practices of carrying out and managing requirements change?" and "what are the challenges faced by practitioners in managing requirements?". The goal of these interviews in general was to understand the change management practices and to identify major challenges. However, during the course of this study, an informal change management process was identified which was later explored. The interviews were recorded and transcribed in full for further analysis. The data collection process spanned over 10 months (Feb 2013 to Jan 2014).

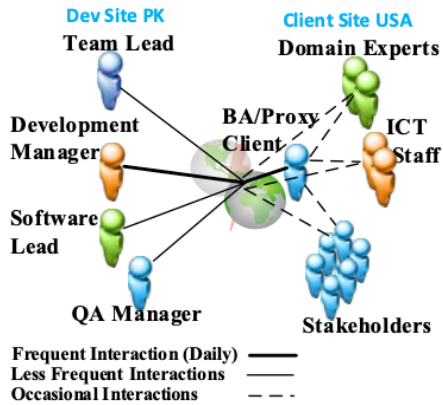

Figure 1. Requirements Collaboration among Stakeholders in WIS Project

Thematic content analysis (TCA) technique [20], was applied to analyze qualitative data collected from the semi-structured interviews. During analysis the data was organized, synthesized, evaluated, interpreted and categorized in order to see patterns, identify themes and discover relationships [21]. The identified challenges for managing requirements change, major factors contributing to these challenges and their implications were placed under appropriate categories that emerged from thematic analysis. Other emergent themes related to actual practices, including InfRC were identified and explored further. The results of TCA carried out by the first author were reviewed and confirmed by two other co-authors.

**B. On-site Observations**
Observations were made during client and development team meetings, team collaboration over requirements and change related activities. These observations regarding activities, roles, sites and process were mapped using (activity based) process mapping [22] technique to understand the RM activities better and create CM model in practice (Figure 3).

**C. Artefact Analysis**
We inspected and analysed a range of artefacts related to RM process using Artefact analysis techniques [23]. The main artefacts included RM process documentation, requirements specification, design specifications, change related emails between the client, proxy client and the development team, from an online issue tracking tool and Requirements Change Logs (RCLs) containing around one hundred change requests.

**D. Analysis Procedure**
The analysis of the data obtained from these three sources helped to ascertain the actual change management practices of the project team and to identify any discrepancies from the prescribed process (Figure 2). Table 1 describes the steps involved in analysing data collected from these multiple sources.

The existing change management practices of the project identified through our analysis were mapped into a model

TABLE 1 PHASES OF THE ANALYSIS PROCESS

| Phase | Description of the process |
| --- | --- |
| Familiarization with data: | Transcribing data (where necessary), reading and re-reading the data, noting down initial ideas. |
| Generating initial codes: | Coding interesting features of the data in a systematic fashion across the entire data set, collating data relevant to each code |
| Searching for themes: | Collating codes into potential themes, gathering all data relevant to each potential theme |
| Reviewing themes: | Checking if the themes work in relation to the coded extracts (Level 1) and the entire data set (Level 2), generating a thematic 'map' of the analysis. |
| Defining and naming themes: | Ongoing analysis to refine the specifics of each theme, and the overall story the analysis tells, generating clear definitions and names for each theme. |
| Producing the report: | Selection of vivid, compelling extract examples, final analysis of selected extracts, relating back of the analysis to the research question and literature, producing a scholarly report of the analysis. |

(Figure 3), which we call the Change Management Process Model in Practice (CMMiP). This model depicts the lifecycle activities of a requirements change based on the actual practices and sequence of activities observed in the case study. The initial draft of the CMMiP model was shared with the vendor's development manager for verification, and the model was updated based on his feedback.

## 4. RESEARCH SETTING

We studied a software procurement and development project at Sync (a fictitious name invented to secure the anonymity of the company). The project involved an enhancement of a web based information system (WIS). The goal of the project was to integrate the WIS interface with existing online tools to facilitate utilization of the available information and improve user experience. The contract outlined high-level scope, objectives, and deliverables for the project. The vendor therefore had to elicit requirements from the existing system, from the stakeholders (various client groups and general public) and other available resources. A bridging role was deemed necessary to mediate or liaise between the client and the offshore development site especially for RE activities. The CEO of Sync, who was onshore with the client in the USA played this role. A mix of waterfall and evolutionary prototyping [24] was adopted as the methodology for the development of WIS. The use of a plan-driven [Waterfall] methodology instead of an agile approach for this project resulted in stricter change control policies.

The vendor team collaborated with three sets of stakeholders: the client's IT staff, domain experts (DEs) and users (shown as 'Stakeholders' in Figure 1). DEs acted as additional clients who participated in requirements-related activities and performed verification and validation services for software releases. Similarly, the client collaborated with two groups of the stakeholders at vendor organization, the onsite BA/proxy client and offshore development manager as well as team leads. Figure 1 depicts the sites and key roles involved in requirements related collaborative activities.

Sync is a CMMI Level–II certified software development company based in USA with an offshore development team in Pakistan. A CMMI based requirements management model [25] was prescribed for use by the management of the vendor organization (Figure 2). The activities shown in the model are linear which start from gathering and analyzing requirements, which are then signed off by the client. Change management activities follow requirements sign off phase and after completion of those activities requirements traceability is managed. Corrective actions are taken if any inconsistencies are identified in the process.

## 5. FINDINGS

Section A describes the actual RCM process observed in practice and the differences identified between the prescribed and actual practices. Section B describes the informal change management process that runs in parallel with the formal one giving details of the sources of InfRC, the reasons to accommodate them and the consequences to the project.

**A. Inconsistencies with Prescribed Process Model**

During the analysis of the data from the interviews, observations and artifact analysis, several differences were identified between the prescribed model (Figure 2) and the actual requirements management practices (Figure 3). Data analysis also revealed that the prescribed process was not fully followed in the vendor organization. The main lifecycle activities of a requirement/change observed in actual practice were: Elicitation, Analysis (and Negotiation), Model & Design, Detailed Specification, Negotiation and Prioritization, Implementation and Test & Fix Cycle.

The RM process model prescribed for Sync however, did not capture the complex and iterative nature of how requirements were actually managed. The problems identified with the prescribed model are discussed here:

***Prescribed model lacks coverage for RM activities:*** The WIS project methodology involved evolutionary prototyping. Prototyping was also utilized to model, verify, validate, negotiate and prioritize requirements as well as changes during the project. Updated versions of design documents and prototypes were shared with the client to obtain their official approval prior to the development work. None of these activities were captured in the prescribed model (Figure 2).

***Differences in responsibility:*** The participant interviews and other process artefacts revealed that at least five individuals from two different sites contributed to elicitation and analysis activities as opposed to the two roles shown in the prescribed model.

On the other hand, according to the prescribed model, members of both the development and management team were responsible for creating change related documentation. However, the analysis of the RM documentation and the interview data revealed that 12 out of 15 documents were produced by the quality assurance department.

***Differences in sequence of activities:*** Requirements traceability is shown as the fifth activity that starts after the requirements change management. In practice, based on the change process documentation, traceability started once the initial requirements were signed off by the client and baselined.

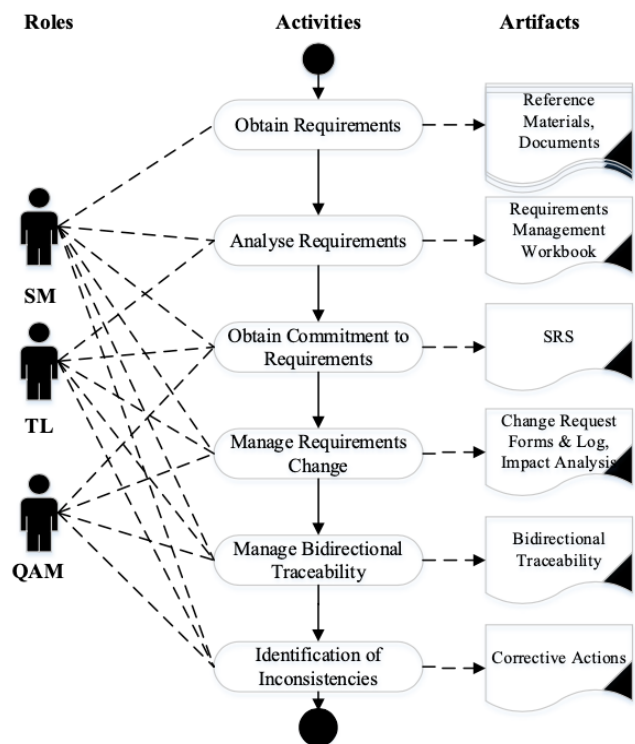

Figure 2. Requirements Management Process Model (Activity Diagram)

Similarly, according to the contract, the design specification sign-off was to be at the end of the requirements elicitation. However, in reality the alpha release testing and feedback were used for elicitation, clarification and modification.

***Prescribed model requires significant project documentation:*** Design documents were collaboratively developed by the client and the vendor during elicitation and analysis activities. As a prescribed company practice, changes proposed during elicitation and analysis should be reflected in the design specification documents. However, six out of the eight participants reported that they did not have enough time to update the design documents with all the changes. Furthermore, according to the prescribed practice changes in the design specification documents were to be formally managed through a Tailoring Request Template (TRT). However most of the changes made to the design documents were not formally approved, recorded or managed through the TRTs.

The prescribed RM policy required all functional and non-functional requirements to be recorded in the RM workbook however only ninety-five requirements (estimated to be 20% of the total) were noted in the RM workbook. It resulted in a disconnect between the RM workbook, the initial high level requirements and the detailed design specifications.

***Prescribed CM process lacks support for informal changes:*** According to the prescribed model (Figure 2), changes in requirements could only be managed after requirements sign off. However significant changes were made to the project requirements from the time of the contract award to the actual requirements sign off. In a procurement model of software development, where only high level requirements are incorporated in contractual documents, a natural process of joint requirements understanding and evolution follows. It leads to modifications in existing requirements and scope. In our case study many of these changes were informally requested and accommodated into the existing scope a) without maintaining any record of such changes and b) without invoking the formal change management process.

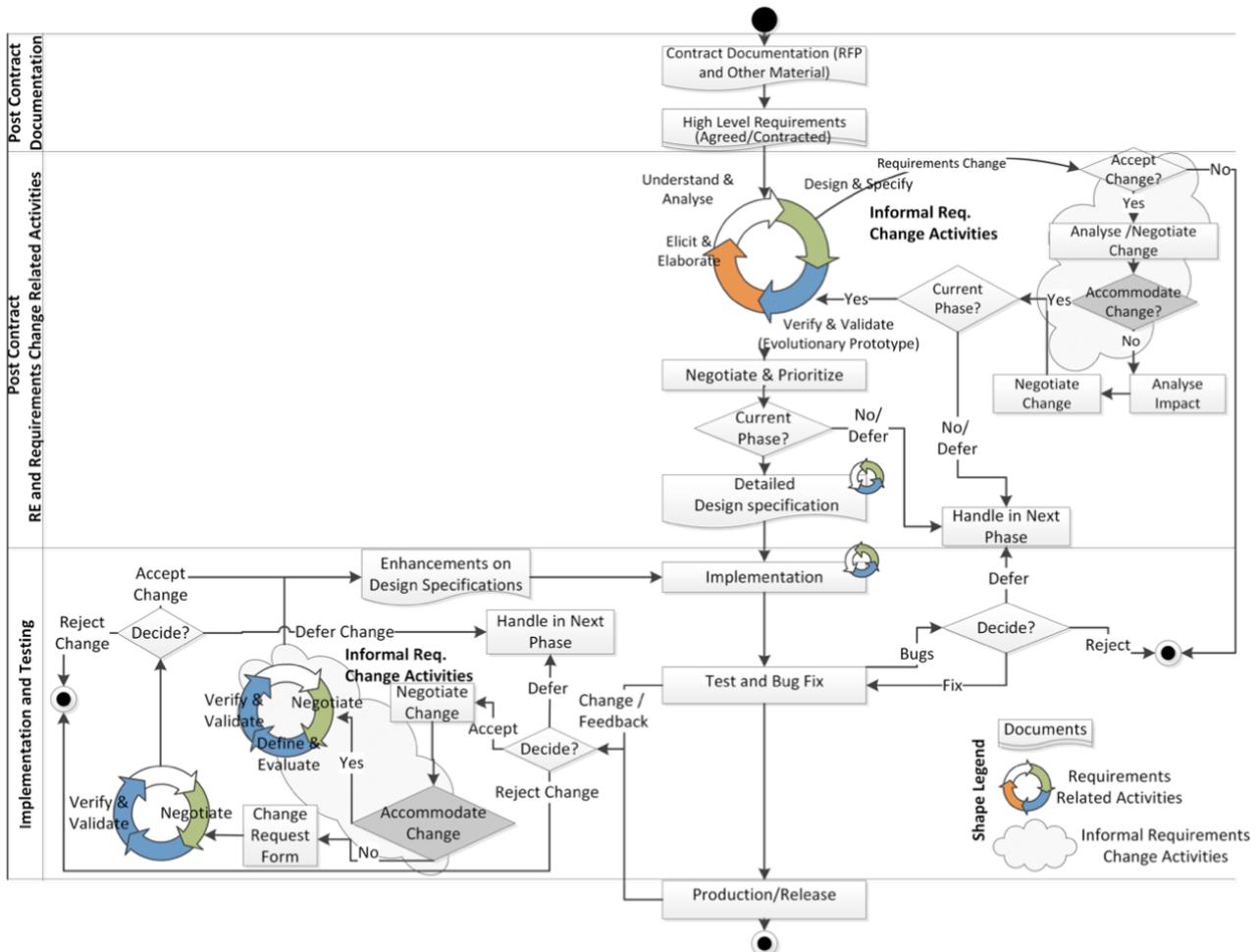

Figure 3. Change Management Model in Practice (CMMiP)

In case of informal change accommodations, no documents (such as change request forms) were produced. Similarly, the client was not billed for the additional effort required for implementing such changes. In such cases the actual practice differed from the prescribed practice which was not captured by the prescribed model (Figure 2).

**B. Informal Requirements Change – A Reality in the WIS Project**

A key finding from our analysis of the actual change management process is the emergence of *Informal Requirements Change* (InfRC). The practices to implement and manage InfRC emerged when participant interview data, the prescribed RM model and change related artefacts were analysed. The analysis helped in ascertaining the actual change management practices and identifying any discrepancies from the prescribed process. This information was mapped out and codified to empirically construct the *'Change Management Process Model' (CMMiP)* in practice shown in Figure 3.

The model in Figure 3 depicts informal requirements change management activities within the two clouded regions. On the right hand side of the main lifecycle activities, informal change activities are depicted (in the clouded region). These activities take place prior to the design specification document signoff. If the change is accommodated by the vendor it returns to the requirements lifecycle activities circle otherwise it undergoes a negotiation cycle before being reprioritized or deferred by agreement. On the left hand side of main lifecycle activities again show the activities for accommodating change as previously noted (in the clouded region). Since many contractual requirements upon elaboration became changes, they passed through the informal activities' cloud (top right of Figure 3).

Changes in requirements continued from elicitation through to implementation and deployment. Accordingly, the decisions to treat changes in contractual requirements formally or informally were also taken throughout the project lifecycle. Therefore, even some of the changes identified during testing and release went through the informal change management activities (bottom left clouded region of Figure 3).

From the vendor's perspective, changes in requirements identified which the vendor considered as out of (contractual) scope, had to be negotiated with the client.

This negotiation resulted in a decision to either include or exclude the change in requirement in the existing scope. To proceed with the contract, the choices available for the vendor were either to get the client's approval to reprioritize requirements, convince the client to adjust the existing cost and project schedule or bear the cost of these changes. In some cases, the vendor agreed to accommodate such change requests, which they considered outside the initially agreed contractual project scope. Such accommodations of changes in requirements were treated informally and were carried out without charging the client or invoking the formal change management process. According to the development manager none of the changes identified during elicitation, design and specification period were considered formal, he stated

*"Changes that come during the requirements or design phase are not considered 'changes', they are better understandings of (the same) requirements. That is why we do not put those changes into our formal change management process"*

Similarly, changes in requirements requested by the proxy client (CEO of the vendor company) were handled informally and implemented without (officially) adding extra time and effort to the existing project plan.

**C. Sources of Informal Requirements Change (InfRC)**
Several sources of InfRCs were identified in this case study. We also observed an imbalance of power relationship between the development team and the proxy client, as discussed below.

***Imbalance of power relationship between the development team and the proxy client.*** One of the main reasons for the project going through many informal change implementations was the role of the proxy client. There was an asymmetrical power balance between the proxy client and the development team [26]. Having the domain knowledge and familiarity with the client's culture and language earned the proxy client respect and gave him a sense of power over the members of the offshore development team. The development team members often relied on his domain knowledge, discernment and comprehension for verification and clarification purposes of their understanding of requirements.

The position of the proxy client as a CEO of the company afforded him the advantage of suggesting and having informal requirements implemented. The power difference made it difficult for the development team members to say 'no' to his informal change requests. One of the managers discussed how these informal requirements kept coming in and getting changed by the proxy client almost on a daily basis.

*"The proxy client says that we would build graphs to present the data, and the next day we are told to create a certain type of graph and the following day the proxy client would say no develop a 3D graph. So almost every second or third day the requirements are changed."*

He continued to explain that such frequent informal changes were not even considered changes by the proxy client.

*"He [proxy client] does not even consider those as requirements change...he says there is no harm in tweaking the UI."*

The implementation of these informal change requests (in a 'timely' manner) was also a cause of contention between the development team and the proxy client. Often the proxy client would argue with the team members and ask why "this sort of small change has not been done in two weeks?"

***Changes with low implementation effort:*** Another source of informal changes was the modifications to the requirements that required low implementation effort. These changes were often identified during elicitation and design phase but had a 'minor' impact on cost and schedule. These were not considered as formal changes by the vendor and hence were treated informally often bypassing the complexities or regulations of the formal change management process that required involving the CCB and getting change request approvals). The development manager reported that such changes were accommodated within the existing scope of the project without charging the client.

*"We (the client and the vendor) have some initial understanding of the requirements scope and budget. If we see that there is a small change /difference [in scope] which is absorbable by the initially proposed budget to the client then we implement the change on our end with no additional cost to the client".*

If the change demanded a significant effort based on an informal evaluation, it was renegotiated with the client

*"If we see that it is going outrageous and we need much more effort and resources then we let the customer know... it then becomes a business decision which takes its own [formal or informal] path."*

***Requirements with subjective nature:*** The project involved development of a web interface with better presentation and improved user experience (UX). The exact definition of the UX remained elusive as it was not clearly defined. Given the subjective nature of the desired UX often many requirements in the studied case had to be 'invented'. One of the software developers explained:

*"The project we have is more of a product than a project in which we have to innovate the requirements. The customer has simply given us the data that we need to present it in a useful, appealing and better way."*

The conception, formulation and discovery of the hidden user needs came through inventing new product features using innovative technical solutions.

**D. Reasons to Accommodate Informal Change Requests**
In the case of informal change, the cost of implementation was not charged to the client. The development team members provided various reasons for accommodating informal changes and managing them informally.

***Low Implementation Effort***: If the team members perceived the change to be implemented required 'low' effort (up to five hours) it was accommodated without charging the client.

*"So if it is one to four hours' work, we tend to accommodate it."*

**UI Changes**: UI changes were considered 'low' effort and were implemented without invoking any formal process.

*"The changes which come within UI, we do not consider them changes at all... the changes to the workflow are a part of the change management process."*

**Business Relationship & Goodwill**: Change requests made by the client (considered low in impact), especially during elicitation and design phases were often implemented without any cost to the client. This was meant to establish good client relationship and to retain the client.

*"The owner of the company who is directly dealing with the client, takes that decision [to accommodate change]...obviously it is based on the client relationship"*
*"to facilitate the customer...to create some goodwill... [and] yes to retain the client."*

**Nature of the Client**: Changes were also accommodated if the client was perceived to be 'stubborn'.

*"It [the response] also depends on the client, you know some clients are good they quickly understand [the impact and effort] and tell you that they will pay for that. Others are a bit stubborn so you have to accommodate them."*

**Peer Pressure & Internal Threshold to Save Documentation Effort**: The vendor's offshore team was also faced with an implicit internal threshold about when to use a change request form. The development team members were 'encouraged' to simply carryout 'small' development tasks that involved a development effort of four to five hours.

*"For minor work we make the CRF but we don't bill it. ... If the change is just a four hours' work, we do not bother sending a CRF to the client... sometimes we are [informally] told to just implement it."*

*"See first there has to be a consensus from our CCB members, to see if it really is a change [worth going through the process of CRF]."*

### E. Implications of Informal Requirements Change Management

Several implications of informally managing requirements changes were identified in the studied project. These implications included: added pressure on the team to meet deadlines, extra time and (uncompensated) effort, delay in release dates, misunderstanding and conflicts among the team members, problems in change understanding, confusion during testing and lack of requirements traceability.

Continued accommodation of informal change requests from the client were portrayed as a business strategy, a relationship building activity and an internal mechanism for efficiency. Similarly, InfRCs received from the proxy client were hard to refuse for the development team.

**Added pressure on the team**: the added impact of even 'low effort' change requests put pressure on the development team.

*"Yes these [informal] changes put pressure on the team especially when they come at the tail end of the project"*.

**Delivery date delays**: the required extra effort often contributed to delays in delivery.

*"We have to put in the extra time ... if the change was of twelve hours and the client asked us to it in six 6 hours, and we agreed, then obviously the delivery deadlines are extended"*.

**Miscommunication, misunderstanding and conflicts**: handling informal changes also contributed to team issues including miscommunication, misunderstandings and conflicts.

*"The first and foremost problem is that we don't know that a change has arrived. The proxy client discusses and decides a change with the development manager who (casually and temporarily) writes it down on Notepad. When the developer comes next day he is asked to implement that change. The change gets implemented and the QA department is not even aware of that change."*

**Traceability related challenges**: Our artifact analysis revealed that, for informal change requests, often changes were carried out in the code itself, but design specifications, use cases and test cases were not updated. This practice resulted in breaking the traceability links between the aforementioned artefacts. In the interviews, the quality assurance department also spoke of the lack of traceability.

The QA Manager noted that the constant influx of informal changes, especially related to user interface, prevented the appropriate traceability matrix updates.

*"We do not or cannot develop or maintain a traceability matrix because we get bombarded with many requirement...generally if it is UI change, the traceability coverage is limited."*

**Testing related challenges**: Testing informal change implementation was quite challenging in the absence of critical information such as what and where the changes were made.

*The tester gets to know about the change only when he looks at the actual screen itself and goes like 'why is this screen appearing or behaving different?' Then he realizes that the screen was changed as a result of the last night's meeting."*

Due to lack of traceability, the testing team was often unaware of the informal requirement changes. As per participant interviews, these changes, therefore, were often reported as bugs during testing. This not only caused tensions between the development and testing team members, it also required significant amount of coordination overhead and testing rework to resolve these issues.

## 6. DISCUSSION

In this section we discuss the inevitability of InfRC, the relatively limited coverage of InfRC in the literature, why InfRC appears to be a pervasive phenomenon, why

| Number | Project | Title | PM (IEC) | PM (IEPL) | Lead | Dev | QC QA | Total | Date Requested | Requested By | Date Approved /Rejected | Status | Deployment Status | Acceptance Date | Billing Status | Comments |
|---|---|---|---|---|---|---|---|---|---|---|---|---|---|---|---|---|
| CR00027 | *** | Total Innerwork Page Count Enhancement | 1 | 1 | 2 | 18 | 2 | 24 | 27-Jun-2009 | TSG | 2-Jul-2009 | Approved | On Production | 2009 | Billed | |
| CR00029 | *** | Ticket# 2545 - Placing an onWait order to onHold | 3 | 2 | 2 | 23 | 10 | 40 | 22-Jul-2009 | TSG | 29-Jul-2009 | Deffered | - | - | NA | |
| CR00039 | *** | Synchronize SGGS_ORDER_ITEM_DETAIL | 0 | 1 | | 5 | 4 | 10 | 23-Nov-2009 | TSG | | Approved | On Production | 2010 | Billed | |
| CR00043 | *** | Ticket# 3168 Prefix problem in order file names | 1 | 1 | 1 | 8 | 3 | 14 | 29-Sep-2009 | TSG | 4-Aug-2010 | Rejected | - | - | NA | |
| CR00049 | *** | Introduce Parent Child Relationship between publishers | 4 | 9 | 9 | 44 | 24 | 90 | 30-Sep-2010 | TSG | 18-Nov-2010 | Approved | On Development | | | This work was completed. *** changed requirements. This will be again done as a new CRF00056 under ***. |
| CR00056 | *** | Introduce Publisher Grouping | 4 | 9 | 3 | 96 | 24 | 136 | 24-Feb-2011 | TSG | 2-Jun-2011 | Approved | On Development | | | |

Figure 4 Requirements Change Log (RCL) Version 2

developers accommodate InfRC and the implications for researchers and practitioners.

**A. Inevitability of InfRC**

Project requirements may never be complete, in fact sometimes they are purposefully left incomplete, which leaves room for multiple interpretations and change [27]. An effective RE process must deal with situations where formalized description of both functional and nonfunctional requirements may not be available [27]. In software procurement and development projects "open-target requirements" are recommended since it is hard to specify requirements in detail upfront [28]. In such context the customers specify their demands and expectations and the vendor responds with how they can meet the demands. Again, this approach leaves plenty of room for requirements elaboration and change, some of which may end up being informal change requests. As Wiegers states [8], implementing requirements change is not free. The recommended strategy to cope with (the 'forced' and unwanted) informal changes according to Weinberg [29], is to say 'no'. However, in the studied case the powerful role of the proxy client did not allow the development team members to have the option to say 'No'.

Although a formal CMMI-based change management process was in place, informal changes requested by the proxy client bypassed this process. Furthermore, these informal changes were implemented within the same time and scope. This scope creep could have resulted in many unwanted outcomes of blown out project cost and missed deadlines. The project however was kept on course for an on-time delivery at the expense of the development team's unrewarded extra work. Fearing admonishment by the proxy client or even worse, losing their jobs, the team members did not say 'no' to these informal requirements. This could also have been because "people done like to say 'no'," and development teams can receive intense pressure to always say "yes" as noted in [8].

In the studied case, most of the informal changes in requirements requested by the proxy client were treated as 'emergency fixes' which had to be carried out immediately. This meant that almost all formal change management steps had to be bypassed. Harjani and Queille [17] consider change requests that bypass certain formal steps as *variants* of the formal or instantiated process. These fixes however are used to prevent a disaster or to modify software urgently. Time constraints on these changes make them incompatible for a formal process of maintenance hence a short procedure becomes necessary. None of these were applicable for the informal change requests made by the proxy client in the studied project.

Describing the minimal steps involved, Harjani and Queille report that, when emergency changes are deemed necessary (which was almost always the case in this case study) only a minimal solution and impact analysis is carried outdone by the most experienced staff followed by change implementation and testing.

Sommerville [16], suggests that implementing changes quickly and directly into the system without following a formal change management process adversely affects the system. Since changes are made directly to program code without modifying the requirements or design; the design and code become inconsistent. Furthermore, in situations where a quick and workable solution is chosen instead of the best solution, it accelerates software ageing [16]. As noted in section V, not communicating informal changes resulted in traceability issues and confusion among stakeholders.

Sabaliauskaite et al. [30] note that when requirements engineers do not inform developers and testers about changes in requirements, it creates several challenges for testing. Testing teams face extreme difficulties to identify the right people who have change related information or developers who have implemented the changes. Since change related information is not updated testers are not aware about the changes that have occurred. It further causes traceability and requirements verification to be a big issue. Similarly Bjarnason et al. colleagues [31] note that lack of change communication between requirements engineering and testers contribute to wasted effort and frustration as well as lack of motivation to work. Furthermore it often results in quality issues for the software output in terms of meeting client expectations [31]. Surprisingly, even from the client side, some of the formalized and accepted changes were implemented without any cost to the client (See CRF00049 and CRF00056, Figure 4). The possible reasons could be any of the previously stated rationale such as creating goodwill, nature the client or building relationship as a business strategy as noted in Section V.

**B. InfRC and Formal Change Models**
The focus on formal RCM that is prevalent in the literature [16] has meant that informal RC has received very limited treatment. Requirements change in general is often viewed negatively [32]. Handling of informal changes is similarly viewed as a negative activity infamously associated with "scope creep", "Gold plating", creeping elegance [5, 8] and "skunkworks activity" [7]. Approaches to its treatment therefore appear to have been overlooked, apart from the agile movement and its focus on permitting and even encouraging change [33]. While agile processes address a change orientation they typically have some notion of a baseline, and apply practices of time- boxing and prioritization as the primary RCM mechanisms [34].

So literature shows a contrast between the traditional and agile approaches, with the former focused on task and control and the latter on people and practices. In the classic requirements change management models [15] formal processes driven by CCBs and CRFs is a typical approach and ignores informal processes for dealing with requirements change. So why are the findings in this study different? For instance, the model presented by Niazi et al. [15] was extracted solely from interview data and their interpretation of how the studied organizations managed requirements changes. In contrast our study draws upon multiple empirical data sources and methods which are mutually supporting: interview data, electronic artefacts, process mapping and close observations of practices based on significant periods in the field setting. This closeness to practice enabled us to see the divergence between practitioners' actions and the classical models of requirements change.

**C. InfRC and Evolving Practices for Managing Change**
Previously [32] we discussed the role of spreadsheets in the evolution of requirements management practices in a Global Software Engineering setting. It was evident from tracking the requirements change logs over time (6 years) that the team had developed increasing sophistication in managing and in effect formalising InfRC, by recording time spent by developers on changes which were not billed to the client. We can see from Figure 4 that the development team's practices evolved to introduce effort hour estimates invested in implementing InfRC. These estimates were not present in the earlier version of their Requirements Change Log. So it is apparent that the practitioners were not only aware of the implications of InfRC but they had also started to develop some mechanisms for coping with them.

## 7. LIMITATIONS

The findings in this study are drawn from a single case and while our observations and synthesis with the previously published research suggest its wider relevance, subsequent studies would be needed to demonstrate the validity of the phenomenon and the applicability of the model in other settings. The researchers had limited access to the end client which resulted in our deriving these finding mostly from the vendor's perspective. The client perspective was able to be seen through the vendor role acting as the proxy client. A more direct end client interaction in the offshore site may have provided additional insights. Furthermore, in this study we do not claim generalizability as the reported single case may not be representative of all similar contexts. Nevertheless InfRC resonates with our own observations over many years in practice. The phenomenon of informal change may be explored better by using multiple case study in different organization having different cultures; which is our plan for future work.

## 8. IMPLICATIONS FOR RESEARCH

In this study the role of proxy client appeared to generate a large amount of InfRC whether that is specific to this case or a more general deficiency in a proxy client's boundary spanning role would be a fruitful area for research.

In addition, better understanding of InfRC and its applicability in software development projects in other domains needs further investigation. We have argued that agile methodologies largely deal with InfRC through prioritization mechanisms and by time boxing but we believe that InfRC still presents challenges in agile projects and warrants closer investigation.

The market driven software aspects of this study as covered in [33], have highlighted challenges which seem to drive InfRC. It is necessary to find a good trade-off between requirements corresponding to perceived user needs and new, invented ones that may provide a competitive advantage through ground- breaking technology. Finding a good balance between technology-driven and needs-driven requirements may be a delicate challenge. Research into market driven software development approaches and management of InfRC could help address these challenges.

More generally it appears that requirements change as a phenomenon is both under theorised, and poses challenges in practice. The refined understanding developed in this study could lead to a deeper study with a focus on developing a broader theory of change in software, possibly akin to that proposed in an organizational context by [35]. Van De Ven and Poole argue that a pluralist approach that combines a rich array of possibilities to study change in an organization may provide richer insights into the phenomenon of change [35].

## 9. CONCLUSIONS

In this paper we have investigated the phenomenon of informal requirements change (InfRC). We have analyzed the change management practices of a geographically distributed development team and have derived a novel process model for both formal and informal requirements change management. The model reflects upon a procurement and software development context from a vendor perspective. In this setting InfRC represents an under reported and variably managed project dimension of hidden work which demands suitable buffers to manage informal change activities in development projects. Handling of InfRC contrasts with the classical perspective of requirements change management, which involves highly formalized and rigorous change processes. Moreover, in change oriented agile development settings we believe that InfRC is an under researched phenomenon.

We argue that InfRC is endemic to software development and imposes pressures on development teams. We need to recognize that informal requirements change serves a necessary and useful purpose, rather than simply being a product of poor development practice and weak project management. Thus we need further research to better understand this phenomenon and develop strategies and techniques to accommodate its prevalence in practice.